\documentclass[twocolumn,preprintnumbers,amsmath,amssymb,prl]{revtex4-1}
\usepackage{amsmath}    
\usepackage{graphicx} 
\usepackage{kotex}      
\usepackage{dcolumn}
\usepackage{hyperref}
\usepackage[toc,page]{appendix}
\usepackage{multirow}
\usepackage{subfig}

\begin{document}    

\title{Motion of a Test Particle in the Reissner-Nordstr{\" o}m Spacetime}    

\author{Moonju Hong}	

\affiliation{Dept. of Physics, Pohang University of Science and Technology, Pohang 790-784, Korea \\
E-mail: moonj0928@postech.ac.kr \\
September 26, 2017}
\begin{abstract}
\section{abstract}
This paper focuses on the motion of a test particle moving around the Reissner-Nordstr{\" o}m black hole. It deals with circular motion and radial motion of the neutral massive test particles, and shortly handles circular motion of the charged massive test particles. Both neutral and charged particles are affected by black hole's charge, but it is due to the fact that charge of the black hole bends the spacetime more strongly. This procedure has nothing to do with electromagnetic interactions, and these are only considered for charged test particles. However, it only treats mathematically easy, approximated situations and general motions and complex motions will not be discussed. This paper has tried to get some physical information only with the easiest mathematical tools and without difficult concepts that general relativity contains. Contents of this paper would be suitable for those who want to know something about the Reissner-Nordstr{\" o}m black hole, but does not have much knowledge in this field. They can begin their intellectual  journey with this paper.

\end{abstract}
\maketitle	

\section{Introduction}
Reissner-Nordstr{\" o}m spacetime is one of vacuum solutions of Einstein equation. Although this solution is generally considered unphysical since it is static and charged. In the scale of stars, this is hardly probable because particles accumulated to create the black hole have their own angular momentum and charges will attract the opposite charges until the black hole becomes neutral. This spacetime, however, can be treated with a bit simple mathematical approach compared to Kerr or Kerr-Newman spacetimes and might suggest us some physical information. In this paper we will see how test particles moving in the Reissner-Nordstr{\" o}m spacetime behave in the most easiest cases: circular motion and radial motion. \\
To begin with, Reissner-Nordstr{\" o}m metric has this form:
\begin{equation*} \tag{1}
ds^2 = -(1-\frac{r_S}{r}+\frac{r_Q ^2}{r^2})c^2 dt^2 + (1-\frac{r_S}{r}+\frac{r_Q ^2}{r^2})^{-1}dr^2 + r^2 d\Omega ^2. 
\end{equation*}
where $r_S$ is proportional to the mass of the black hole and $r_Q$ is proportional to the net charge of the black hole.\cite{metric} We will ignore an additional term given by magnetic charges throughout this paper(letting $P=0$.) This metric form ensures its spherical symmetry, and tells that the spacetime will depend on the net charge and mass of a star. \\
Because of its peculiar properties caused by electromagnetic interactions, Reissner-Nordstr{\" o}m black hole shows quite different features from Schwarzschild or Kerr black holes. One of them will be discussed here and it is depicted in FIG. 3. There have been some approaches to gain some physical results from this kind of black hole, since its static property makes it an easy choice to start an intellectual journey. For example, see Ruffini (2005)\cite{rufin}

\section{I. Neutral particle with circular orbit on equatorial plane}
In this case, since the particle is neutral, there is no electromagnetic interaction between the test particle and the Reissner-Nordstr{\" o}m black hole. Spherical symmetry of the Reissner-Nordstr{\" o}m metric ensures that the motion will remain on the equatorial plane, thereby described by $(t, r, \frac{\pi}{2}, \phi)$ with $\dot{\theta} = 0$. \\
With these initial conditions, geodesic equation for $\mu = 0$ yields
$$ \frac{d^2 t}{d\tau ^2} +2 \Gamma^0 _{10} \frac{dt}{d\tau} \frac{dr}{d\tau} = 0, $$
simplified into
\begin{equation*} \tag{2}
 \frac{d}{ds} [(1-\frac{r_S}{r}+\frac{r_Q ^2}{r^2})\dot{t}] =0 \,\,\,\,  (\mu = 0) .
\end{equation*}
This result denotes the conserved energy for massless particles, or the conserved energy per unit mass for massive particles. \cite{Killing}
\begin{equation*} \tag{3}
 (1-\frac{r_S}{r}+\frac{r_Q ^2}{r^2})\dot{t} = E = const.
\end{equation*}
 For $\mu = 3$, the equation gives
$$ \ddot{\phi} + \frac{2}{r}\dot{r}\dot{\phi} = 0, $$
which becomes
\begin{equation*} \tag{4}
 \frac{d}{ds} [r^2 \dot{\phi}] =0 \,\,\,\,  (\mu = 3) .
\end{equation*}
In the same reasoning, this result states the conserved angular momentum(for massless particles) or the conserved angular momentum per unit mass(for massive particles.)
\begin{equation*} \tag{5}
r^2 \dot{\phi} = L =const.  
\end{equation*}
Instead of trying $\mu = 1$ equation, we use the Reissner-Nordstr{\" o}m metric
\begin{equation*} \tag{1}
ds^2 = -(1-\frac{r_S}{r}+\frac{r_Q ^2}{r^2})c^2 dt^2 + (1-\frac{r_S}{r}+\frac{r_Q ^2}{r^2})^{-1}dr^2 + r^2 d\Omega ^2. 
\end{equation*}
Dividing both sides by $ds^2$ and substituting Eqs. (3) and (5), for the massive test particles, we obtain
$$ 1 = (1-\frac{r_S}{r}+\frac{r_Q ^2}{r^2})E^2 - \frac{L^2}{c^2 r^4}(1-\frac{r_S}{r}+\frac{r_Q ^2}{r^2})^{-1}(\frac{dr}{d\phi})^2 - \frac{a^2}{c^2 r^2}.  $$
Rearranging terms following the similar procedure presented by L. Ryder\cite{ryder}, we finally arrive at the equation:
\begin{equation*} \tag{6}
\frac{d^2}{d\phi ^2}(\frac{1}{r}) + \frac{1}{r} = \frac{r_S c^2}{2L^2} - \frac{r_Q^2 c^2}{rL^2} + \frac{3r_S}{2r^2} - \frac{2r_Q ^2}{r^3}. 
\end{equation*}
This is the equation that describes the motion of the massive and neutral test particle on the equatorial plane of the Reissner-Nordstr{\" o}m black hole. 
\subsection{Stable, and Circular Orbit Motion}
For further discussion, it would be difficult to handle all kinds of possible motions of these test particles on the equatorial plane. Thus, we focus on a specific case, the circular orbit of the neutral and massive test particle. To find a stable circular orbit radius we start from the metric equation, Eq. (1). Rearranging terms to represent the conserved energy part, $E$, on the RHS and the other terms on the LHS, it becomes as follows:
\begin{equation*} \tag{7}
\frac{1}{2}\dot{r}^2 +\frac{1}{2} (1-\frac{r_S}{r}+\frac{r_Q ^2}{r^2})(\frac{L^2}{r^2}+1) = \frac{1}{2}E^2 \equiv \epsilon.
\end{equation*}
This equation seems like the equation for conservation of energy. In this analogy, the potential energy term for the circular motion can be obtained.
$$ V(r) = \frac{1}{2} (1-\frac{r_S}{r}+\frac{r_Q ^2}{r^2})(\frac{L^2}{r^2}+1). $$
Let the circular orbit radius be $r_C$. Then for the orbit to be circular, this potential should satisfy the condition:
$$ \frac{dV}{dr} \mid _{r=r_C} = 0 .$$
Satisfying this restriction yields a polynomial equation of $r_C$.
\begin{equation*} \tag{8}
r_S - \frac{(2r_Q^2 + L^2)}{r_C} +\frac{3r_S L^2}{r_C ^2} - \frac{4r_Q^2 L^2}{r_C ^3} = 0.
\end{equation*}
If the test particle were massless, the first two terms of this equation will be removed, thereby resulting in the radii of
$$ r_C = \frac{3r_S \pm \sqrt{9r_S ^2 - 32r_Q ^2}}{4}. $$
Especially for the extremal Reissner-Nordstr{\" o}m black hole, where $r_S = 2r_Q$, we have simple radius of
$$ r_C = 2r_Q \,\,\,\,\, or \,\,\,\,\, r_Q, $$
with $r_Q$ being the hozion.\\
Unlike this massless case, Eq. (8) for massive particles will be cubic equation. To make things simple, we first try the special case $ r_C \gg r_Q$ and then see whether it was a reasonable approach afterwards. \\
In this limiting case, the last term in Eq. (8) vanishes, so we solve this quadratic equation:
$$ r_S r_C - 2(r_Q^2 + L^2)r_C +3r_S L^2 = 0. $$
Solving this quadratic equation gives one inner, unstable orbit and another farther, stable orbit. The innermost stable orbit is, therefore, achieved when two orbits coincide, i.e. when the discriminant of this quadratic equation is zero. Again, supposing $r_S \gg r_Q$, the circular orbit radius is given by
\begin{equation*} \tag{9}
r_C \approx 3r_S - \frac{r_Q^2}{r_S}. 
\end{equation*}
\begin{figure*}[htbp]
\subfloat[]{\includegraphics[height=1.8in, width=0.33\textwidth]{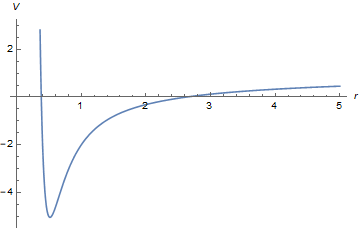}}
\subfloat[]{\includegraphics[height=1.8in, width=0.33\textwidth]{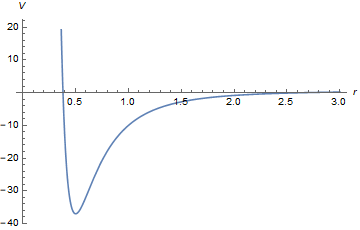}}
\subfloat[]{\includegraphics[height=1.8in, width=0.33\textwidth]{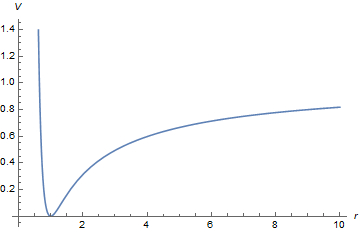}} 
\caption{Graphs of the potential $ V(r) = \frac{1}{2} (1-\frac{r_S}{r}+\frac{r_Q ^2}{r^2})(\frac{L^2}{r^2}+1) $ depending on variables $r_S$, $r_Q$, and $L$, keeping the condition $r_S \geq 2r_Q$ . Whatever values we choose, the genaral shape of the potential does not change. The potential shape resembles that of atoms. Stable circular motion is achieved at the pit. Note that the axes values are different. (a) $r_S = 3, r_Q = 1, L = 1$, (b) $r_S = 3, r_Q = 1, L = 3$, (c) $r_S = 2, r_Q = 1, L = 1$.}
\end{figure*}
This result is compatible with the former constriction $r_C > r_S \gg r_Q$. Thus, we found the radius of stable circular orbit of the massive particle, moving around the weakly charged Reissner-Nordstr{\" o}m black hole. Compared with the Schwarzschild black hole, the stable circular orbit around the Reissner-Nordstr{\" o}m black hole moves a liitle inwards, towards the horizon. Also this result reproduces the Schwarzschild stable orbit, $r = 3r_S$, in the limit $r_Q \rightarrow 0$, or more correctly, when $Q \rightarrow 0$, as desired. It complys with the more precisely calculated radius, provided by Praloy, Ripon, and Subir\cite{stable}, 
$$r_C = 3r_S -\frac{3r_Q^2}{r_S},$$ 
within some coefficients which might depend on the approximation procedure. In case of strongly charged Reissner-Nordstr{\" o}m black hole, we cannot apply the same approximation so one has to solve Eq. (8) directly. Since the potential has the general shape described in FIG. 1, there is one stable circular orbit according to the given potential, and the minimum radius is given by Eq. (9). In the regime that has been used to get Eq. (9), there is another solution; however, it is inside the horizon. Like this, other possible solutions are not physical.

\subsection{Precession Motion}
We finish this section with discussing the precession motion of an orbit. It might not be circular, in general, but elliptical more probably. Before thinking about the precession motion in general relativistic way, we first consider the familiar Newtonian elliptical orbit. It is well known that in Newtonain language,
$$ \frac{d^2}{d\phi ^2}(\frac{1}{r}) + \frac{1}{r} = \frac{1}{p} = \frac{r_S c^2}{2L^2}. $$
Here, the quantity $p$ is the semi-latus rectum of the ellipse, given by
$$ p = a_0 (1-e^2), $$
with the semi-major axis $a_0$ and the eccentricity $e$. This Newtonian equation has the solution
\begin{equation*} \tag{10}
\frac{1}{r} = \frac{1}{p} (1+e\cos\phi). 
\end{equation*}
Now, we go back to the relativistic motion equation for motions on the equatorial plane,
\begin{equation*} \tag{6}
\frac{d^2}{d\phi ^2}(\frac{1}{r}) + \frac{1}{r} = \frac{r_S c^2}{2L^2} - \frac{r_Q^2 c^2}{rL^2} + \frac{3r_S}{2r^2} - \frac{2r_Q ^2}{r^3}. 
\end{equation*}
The first term on the RHS is that of the Newtonian term. To solve this equation, we substitute Eq. (10) into Eq. (6) and neglect terms of order higher than $e^2$(assuming that $e \ll 1$.) Also, further assuming that $r_S$ and $r_Q$ are on the same order and $[(r_S c)/L]^2 \ll 1$, the equation turns into
\begin{align*}
 \frac{d^2}{d\phi ^2}(\frac{1}{r}) + \frac{1}{r} \approx \frac{r_S c^2}{2L^2} & [1-\frac{r_Q^2 c^2}{L^2}e\cos\phi \\
& +\frac{3r_S^2 c^2}{2L^2}e\cos\phi -\frac{3r_Q^2 r_S^2 c^4}{2L^4}e\cos\phi] .
\end{align*}
In this approximation, the last term can be disregarded. This is the differential equation and we already obtained the soultion of the first part - it is the answer the Newtonian equation yields - and the other terms give the solution containing $\phi \sin \phi$. Thus, the total solution of Eq. (6) with some approximations is this:
\begin{equation*}
\frac{1}{r} = \frac{r_S c^2}{2L^2} (1+e\cos\phi) + \frac{r_S^2 c^4}{2L^4}(\frac{3r_S^2}{2}-r_Q^2)e\phi\sin\phi.
\end{equation*}
Because of the cosmic censorship hypothesis, the condition $r_S \geq 2r_Q$ must hold, not to make naked singularity. So the term $\frac{3r_S^2}{2}-r_Q^2)$ in the above solution is always positive. The approximation $[(r_S c)/L]^2 \ll 1$ enables us to synthesize two trigonometric terms into one:
\begin{equation*} \tag{11}
\frac{1}{r} = \frac{r_S c^2}{2L^2}[ 1+e\cos \{\phi(1- \frac{c^2}{L^2}(\frac{3r_S^2}{2}-r_Q^2))\} ].
\end{equation*}
with some minor terms of order higher than $(r_S / r)^2$ ignored. \\
This solution describes the precesss motion of an elliptical orbit with 
\begin{equation*} \tag{12}
\begin{aligned}
\delta\phi & = 2\pi \frac{c^2}{L^2}(\frac{3r_S^2}{2}-r_Q^2) \\
& = \frac{6\pi c^2 r_S}{a_0 (1-e^2)}[1-\frac{2}{3}(\frac{r_Q}{r_S})^2].
\end{aligned}
\end{equation*}
To see its effect, let's imagine an imaginary star which has the same properties as the Sun, except for one change that this imaginary star has net charge of Q. For the Sun and Mercury, the parameters are given by
$r_S =  2.95325008\times 10^3 m$, $a_0 = 5.7909175\times 10^{10} m$, and $e = 0.20563069$.\cite{Allen} Then the cumulative effect for 100 Earth-year becomes
\begin{align*}
\delta\phi_{100} & \approx 43.03''[1-\frac{2}{3}(\frac{r_Q}{r_S})^2] \\
& = 43.03''[1-\frac{2}{3}(\frac{Q^2}{5.9\times 10^{10}})]. 
\end{align*}
\begin{figure*}[htbp]
\subfloat[$r_Q = 0, \,\,\,\,\,  20\pi$]{\includegraphics[height=1.8in, width=0.33\textwidth]{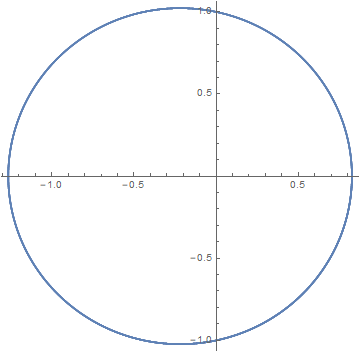}}
\subfloat[$r_Q = 0, \,\,\,\,\, 1,000\pi$]{\includegraphics[height=1.8in, width=0.33\textwidth]{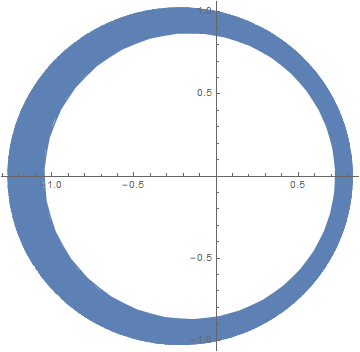}}
\subfloat[$r_Q = 0,\,\,\,\,\,  20,000\pi$]{\includegraphics[height=1.8in, width=0.33\textwidth]{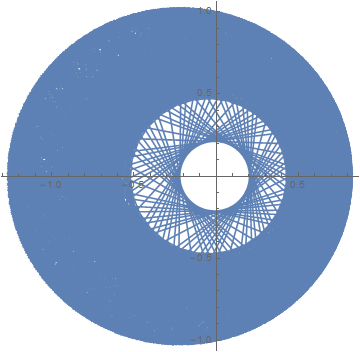}} \\
\subfloat[$r_Q = 0, \,\,\,\,\, 30,000,000\pi$]{\includegraphics[height=1.8in, width=0.5\textwidth]{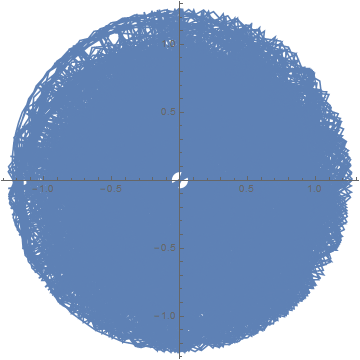}} 
\subfloat[$r_Q = 0.5r_S,\,\,\,\,\,  30,000,000\pi$]{\includegraphics[height=1.8in, width=0.5\textwidth]{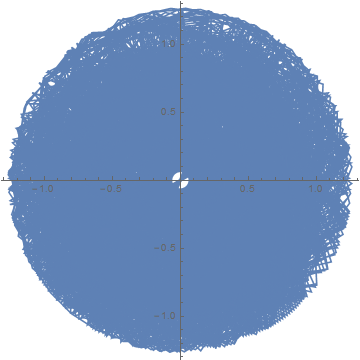}} 
\caption{This figure shows how Mercury revolves the Sun. At the beginning, orbital motion follows the initial elliptical path. As precession effect is accumulated, however, its orbit starts to differ from its original path and covers more space. From (a) to (d), net charge of the Sun is zero ($Q=0$,) as usual. In (e), however, net charge of the Sun is chosen to satisfy extremal condition ($r_S = 2r_Q$). The difference between the neutral Sun and the charged Sun becomes obvious when Mercury revolves more than ten million times.}
\end{figure*}
Just as FIG. 2 shows, the effect of the net charge becomes macroscopic after the planet revolves the star countless times. 
$$ *\,\,\,\,\,\,\,\,\,\,\,\,\,*\,\,\,\,\,\,\,\,\,\,\,\,\,* $$
What these two results - circular orbit change and precession change - tell us is this: Even the neutral test particles are affected by the net charge of the black hole, thus giving different consequnces for Schwarzschild spacetime and  
Reissner-Nordstr{\" o}m spacetime. The presence of charge affects the spacetime and curves it stronger than neutral case when the masses are the same. The more charge it acquires, the more the spacetime curved, like mass does. As expected, not only mass and angular momentum, but also charge is an obvious hair for black holes.\\ \\

\section{II. Neutral Massive Particle with Radial  Motion}
Without loss of generality radial motion of the test particle can be chosen to follow these conditions:
$$ r_0 = R,\,\, \frac{dr}{dt}\mid_{t=0} = 0,\,\, \theta = \frac{\pi}{2},\,\, \phi =0,\,\, \dot{\theta}=\dot{\phi} =0 . $$
Eq. (3) still holds, but angular terms in Eq. (1) vanishes.
\begin{equation*} \tag{13}
ds^2 = -(1-\frac{r_S}{r}+\frac{r_Q ^2}{r^2})c^2 dt^2 + (1-\frac{r_S}{r}+\frac{r_Q ^2}{r^2})^{-1}dr^2 . 
\end{equation*}
Under this situation, we first try to gain t, the time of an observer far away from the origin, as a function of r. Substitute $$ \dot{r} = \frac{dr}{dt}\dot{t}$$
into Eq. (13) and arrange terms, then we have
$$ [c^2 (1-\frac{r_S}{r}+\frac{r_Q ^2}{r^2}) - (1-\frac{r_S}{r}+\frac{r_Q ^2}{r^2})^{-1}(\frac{dr}{dt})^2]\dot{t}^2 = c^2. $$
Applying initial conditions into this equation gives
$$ \frac{dt}{d\tau}\mid_R = \frac{1}{\sqrt{1-\frac{r_S}{R}+\frac{r_Q ^2}{R^2}}}. $$
Since Eq. (3) still valid, we obtain 
$$ E= \sqrt{1-\frac{r_S}{R}+\frac{r_Q ^2}{R^2}} \,\,\,\,\, and \,\,\,\,\, \frac{dt}{d\tau} = \frac{\sqrt{1-\frac{r_S}{R}+\frac{r_Q ^2}{R^2}}}{\sqrt{1-\frac{r_S}{r}+\frac{r_Q ^2}{r^2}}}.$$
With these results, we finally reach at the relationship between the time t and the radial position r.
\begin{equation*} \tag{14}
\begin{aligned}
ct = - & \sqrt{1- \frac{r_S}{R}+\frac{r_Q ^2}{R^2}} \,\,\, \times \\
 & \int_R^r \frac{dx}{(1-\frac{r_S}{x}+\frac{r_Q ^2}{x^2})\sqrt{\frac{r_S}{x}-\frac{r_Q ^2}{x^2}-\frac{r_S}{R}+\frac{r_Q ^2}{R^2}}}. 
\end{aligned}
\end{equation*}
This integral cannot be solved easily, so we again solve it for special case by taking limits $R \rightarrow \infty$(although one can retain R to be finite, it will simplify things) and $ r_S \rightarrow 2r_Q $(extremal approximation,) the integral is solved into
\begin{equation*} \tag{15}
ct = \frac{\sqrt{\frac{2r}{r_Q}-1}}{r-r_Q} . 
\end{equation*}
When the Reissner-Nordstr{\" o}m black hole is extremal, $ r_S = 2r_Q $, the outer event horizon and the inner Cauchy horizon coincide at $r=r_Q$. Therefore,
$$t \rightarrow \infty \,\,\,\,\,\,\,\, as \,\,\,\,\,\,\,\, r \rightarrow r_Q,$$
as desired since in terms of the far away observer, an object falling into the black hole will never cross the horizon. However, it does not make any physical problem because in the view of the infalling observer, it would take finite time to cross the horizon. That is, the proper time integration
\begin{equation*} \tag{16}
c\tau = -\int_R^r \frac{dx}{\sqrt{E^2 - (1-\frac{r_S}{r}+\frac{r_Q ^2}{r^2})}} , 
\end{equation*}
produces finite time even when r is inside the horizon, although it is hard to obtain the general formula. \\
In calculating the proper time, one should note that the denominator of Eq. (16) can become zero, meaning non physical situation when
$$ r = \frac{-r_S + \sqrt{r_S ^2 + 4r_Q^2 (E^2 -1)}}{2(E^2 -1)}. $$
What happens here is well described in Carroll's textbook \cite{carroll}: \\
\begin{figure}[h]
\includegraphics[width=0.5\textwidth]{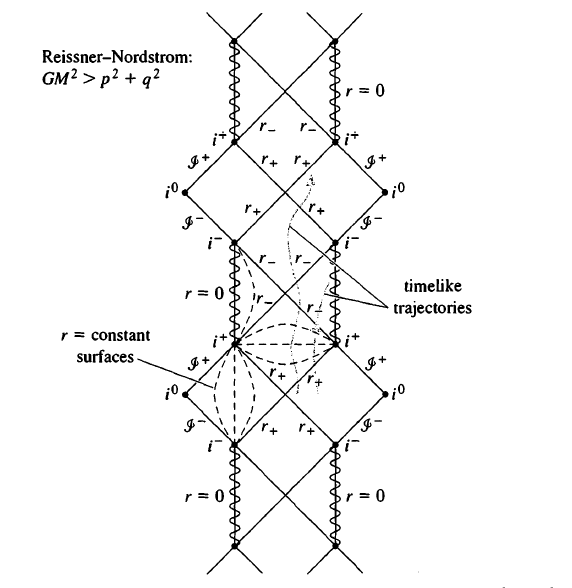}
\caption{Conformal diagram for Reissner-Nordstr{\" o}m black hole. This figure was sampled from S. Carroll's Spacetime and Geometry An Introduction to General Relativity\cite{carroll}. It is shown that an object fallen into the Reissner-Nordstr{\" o}m black hole oscillates back and forth around the horizon. In this paper, $p$, the total magnetic charge is always considered to be zero.}
\end{figure}
After the infalling observer passes the outer horizon $r_+$, he has to pass the inner horizon $r_-$ also. When he crosses the inner horizon, however, r coordinate becomes spacelike, so he can go back to the inner horizon and then cross it from the inside to outside. Then, r coordinate again becomes timelike but since it is reversed, he is forced to move along the increasing r path, thereby passing the outer horizon. Finally, he will be released from the black hole, then again he starts to feel attraction towards the black hole. \\
In this way, the observer can oscillate back and forth around the Reissner-Nordstr{\" o}m black hole's outer horizon. The parodox, however, arises because the far away observer never sees the infalling observer crossing the outer horizon. When the infalling observer crosses the outer horizon from inside to outside, the far away observer will notice that there exist two totally same person: one is still falling, but the other is coming out. To remedy this paradox, it was suggested that when the infalling observer is released from the outer horizon, it will not be the same universe that he lived when he was falling. 
In case of the Scwarzschild black hole, the infalling observer cannot escape but fall into and collide to the singularity. On the other hand, an observer diving into the Reissner-Nordstr{\" o}m black hole can go to other universes as well as escape the black hole. This is another main difference between two types of black holes and it can be implied from the radial motion of the massive particle. \\ \\

\section{III. Charged Particle with Circular Orbit on Equatorial Plane}
We finally deal with the motion of the charged test particle. First, when the particle is far away from the Reissner-Nordstr{\" o}m black hole, the black hole will seem like a point charge with mass. This can be shown easily from Reissner-Nordstr{\" o}m black hole's properties: $ F_{tr} = -F_{rt} = - \frac{r_Q}{r^2}$, else = 0. \cite{wheeler} Electromanetic tensor $F^{\mu\nu}$ is given by
\begin{equation*}
F^{\mu\nu} = 
\begin{pmatrix}
0 & E_r & rE_\theta & r\sin\theta E_\phi \\ 
 & 0 & -rB_\phi & r\sin\theta B_\theta \\ 
 &  & 0 & -r^2 \sin\theta B_r \\ 
 &  &  & 0
\end{pmatrix}
=
\begin{pmatrix}
0 & \frac{r_Q}{r^2} & 0 &0 \\ 
 & 0 & 0 & 0 \\ 
 &  & 0 & 0 \\ 
 &  &  & 0
\end{pmatrix}
\end{equation*}
Noting that $r_Q \sim Q$, we find the expected classical electric field produced by static charge Q. In the neutral particle case, however, the innermost stable circular orbit was constructed at $r_C \approx 3r_S.$ This is close enough to the black hole, so one cannot expect this classical result to be applied. We again follow the procedure done in section I. For stable circular orbit, since radius should remain constant, its derivative will automatically vanish. Then, 
$$ \frac{d^2}{d\phi ^2}\frac{1}{r_C} =0 $$
results in
\begin{equation*} \tag{17}
\begin{aligned}
0 =&  (1-\frac{r_Q qE}{r_S}) - \frac{2r_Q^2}{r_S r_C}(1-q^2+\frac{L^2}{r_Q^2 c^2}) \\ 
& +\frac{3L^2}{c^2 r_C^2} - \frac{4r_Q^2 L^2}{r_S c^2 r_C^3} .
\end{aligned}
\end{equation*}
Also, the fact that stable orbit occurs at the inflection point of the orbital equation requires derivative of Eq. (17) to be zero. These two conditions are written by powers of $r_C$ as
\begin{equation*} \tag{18}
\begin{aligned}
0 = & (1-\Lambda)r_C^3 - \frac{2r_Q^2}{r_S}(1-q^2+\frac{L^2}{r_Q^2 c^2})r_C^2 \\
& + \frac{3L^2}{c^2}r_C - \frac{4r_Q^2L^2}{r_S c^2}.
\end{aligned}
\end{equation*}  
\begin{equation*} \tag{19}
0 = 3(1-\Lambda)r_C^2 - \frac{4r_Q^2}{r_S}(1-q^2)r_C-\frac{4L^2}{r_S c^2}r_C +\frac{3L^2}{c^2}.
\end{equation*}  
where $\Lambda \equiv (r_QqE)/r_S$. We now solve these two equations and suppose that $(\frac{r_Q}{r_S})^2 \ll 1$, then the last term in Eq. (18) vanishes. Therefore, the approximated answer is obtained.
$$ r_C \approx \frac{3}{2}r_S [1+ \sqrt{1- \frac{4r_Q^2}{r_S}(\frac{1-\frac{2}{3}\Lambda - \frac{1}{3}q^2}{1-\Lambda})}]. $$
When the charge of the particle is small enough, this can be roughly stated as
\begin{equation*} \tag{20}
r_C \approx 3r_S - \frac{3r_Q^2}{r_S}(1+\frac{r_Q E}{r_S} q).
\end{equation*}  
This is the exact solution presented by Praloy, et al. in more rigorous way.\cite{ans} Note that the coefficient of $q$ is vanishingly small since we assumed that $(\frac{r_Q}{r_S})^2 \ll 1$. \\
What is interesting here is that when $q=0$, the neutral test particle solution is retrieved. Moreover, this result also tells that the net charge of the particle can be translated into the change of the charge of the Reissner-Nordstr{\" o}m black hole. According to the sign of the test charge, the innermost circular orbit becomes closer or farther. \\
Not just being satisfied with the circular case, one can solve the equation 
\begin{equation*} \tag{21}
\begin{aligned}
\frac{d^2}{d\phi ^2}(\frac{1}{r}) + \frac{1}{r} = & \frac{r_S c^2}{2L^2} - \frac{r_Q^2 c^2}{rL^2} + \frac{3r_S}{2r^2} - \frac{2r_Q ^2}{r^3} \\
& +\frac{r_Q qc^2}{L^2}(\frac{r_Q q}{r} - \frac{1}{2}E)
\end{aligned}
\end{equation*}  
to see the full picture. This is Eq. (6) plus some additional terms due to the electromagnetic interaction of the black hole and the test particle. \\

\section{Conclusion}
This paper has concentrated on the motion of a massive test particle moving near the Reissner-Nordstr{\" o}m black hole. First, we treated neutral test particles. Although they are neutral, so there should be no electomagnetic interactions, it turns out that even neutral test particles show different motions compared to that of the Schwarzschild cases. This is because the presence of charge can curve the spacetime as mass and energy does; however, it should be noted that electromagnetic interaction itself has nothing to do with curvature just as gravity is the curved spacetime itself. When the Schwarzschild black hole has started to get net charge, it becomes Reissner-Nordstr{\" o}m black hole and it begins to curve spacetime more strongly than before. A neutral test particle can follow a cirrcular orbit around the black hole. The innermost stable circular orbit it can have differs from two black holes. For schwarzschild black hole, the radius is only determined by $r_S$, because it is the only hair it has, while the orbit radius around  the Reissner-Nordstr{\" o}m black hole depends on both $r_S$ and $r_Q$ and becomes closer, as expected. \\
Second, we paid attention to the fact that Schwarzschild metric is used to explain precession of the perihelion of the Mercury. From this, we imagined a star which has the same properties with the Sun except for its net charge, and calculated how it affects Mercury. It is described in FIG. 2 and we found that its effect is quite small, so one can notice the difference after Mercury revolved tremendous times - even for the extremal case which has the most drastic effect. \\
After that, neutral test charge's radial motion was considered. Like all other kinds of black holes, a person falling into the Reissner-Nordstr{\" o}m black hole never crosses the event horizon in the view of the far away observer. However, another observer falling together with the infalling person surely sees him taking finite time to pass the horizon. Also, radial motion implies a weird characteristic of the Reissner-Nordstr{\" o}m black hole. Even though it has the name 'black hole,' already fallen person can go outside the event horizon; however, he will be in another universe. This is quite an interesting story, although the Reissner-Nordstr{\" o}m black holes are not realistic objects in the universe. \\
Finally, charged test particle was handled, but it was not discussed in depth since it is difficult to insert electromagnetic interactions into the Reissner-Nordstr{\" o}m metric and develop a theory. It was shown that the existence of net charge of the test particle can be translated into an increased or decreased charge of the black hole. Also, the innermost circular orbit radius approached that of the neutral particle when we let the charge of the particle be zero. \\
Since Reissner-Nordstr{\" o}m black holes distinguish themselves from other kinds of black holes by their charge, most dramatic and amusing effects come from their interaction with the charge of the test particle. Because of its mathematical hardship, however, this paper has satisfied only with the easiest cases. Who one to go further and see what happens could solve Eq. (21). For example, this equation can teach us how electron beams caused by supernovae or whatever behaves when they pass the Reissner-Nordstr{\" o}m black hole nearby. Gravity only works as a convex lens for all kinds of matters, but Reissner-Nordstr{\" o}m black hole might be able to scatter charged particles, working as a concave lens. Or because charged particles emit electromagnetic waves when it is accelerated, its orbit will not remain stable. One can see how it will fall into the Reissner-Nordstr{\" o}m black hole in this way. \\
Although this paper has tried to gain some simple physical results in some simpliest cases, general orbits of the charged test particles' motions are well described in the paper of Grunau and Kagramanova.\cite{orbit} This paper is strongly recommeded for those who want to see particles' exact behaviors not in equations, but in figures. Also, works done by Praloy, et al., which has been cited all along this paper shows how neutral and charged test particles move in much more rigorous way.


\section{Reference}
\begin{thebibliography}{9}
\bibitem{metric}
C. Misner, K. Thorne, and J. Wheeler, Gravitation, 1st edition. (W. H. Freeman and Company, 2000), p. 921
\bibitem{rufin}
R. Ruffin, Charges in gravitational fields: from Fermi, via Hanni-Ruffini-Wheeler, to the "electric Meissner effect", (Unpublished). (2005), 0503439 
\bibitem{Killing} 
S. Carroll,  Spacetime and Geometry An Introduction to General Relativity, new international edition. (Pearson, 2014), pp. 207-208
\bibitem{ryder}
L. Ryder, Introduction to General Relativity, 1st edition. (Cambridge University Press,  2009), pp. 158-160 
\bibitem{stable}
D. Praloy, S. Ripon, and G Subir, Motion of charged particle in Reissner-Nordstr{\" o}m spacetime: A Jacobi metric approach, (Unpublished). (2017), 7, 1609.04577 [gr-qc] 
\bibitem{Allen}
C. Allen, Astrophysical Quantities, 4th edition, editied by A. N. Cox (Springer-Verlag, New
York, 2000) 
\bibitem{carroll}
S. Carroll,  Spacetime and Geometry An Introduction to General Relativity, new international edition. (Pearson, 2014), pp. 257-259
\bibitem{wheeler}
C. Misner, K. Thorne, and J. Wheeler, Gravitation, 1st edition. (W. H. Freeman and Company, 2000), p. 877
\bibitem{ans}
D. Praloy, S. Ripon, and G Subir, Motion of charged particle in Reissner-Nordstr{\" o}m spacetime: A Jacobi metric approach, (Unpublished). (2017), 12, 1609.04577 [gr-qc] 
\bibitem{orbit}
S. Grunau and V. Karamanova, Geodesics of electrically and magnetically charged test particles in the Reissner-Nordstr{\" o}m spacetime: analytical solutions, Physical Review D, (2011). 83(4): doi:10.1103/physrevd.83.044009

\end{thebibliography}
\end{document}